# OVERVIEW OF BEAM–BEAM EFFECTS IN THE TEVATRON*

V. Shiltsev[#] for the Tevatron Beam–Beam Team, FNAL, Batavia, IL, USA

*Abstract*

For almost a quarter of a century the Tevatron proton–antiproton collider was the centrepiece of the world's high-energy physics program, from the start of operation in December 1985 until it was overtaken by the LHC in 2011. The initial design luminosity of the Tevatron was $10^{30}$ cm$^{-2}$ s$^{-1}$; however, as a result of two decades of upgrades, the accelerator has been able to deliver 430-times higher luminosities to each of two high-luminosity experiments, Collider Detector at Fermilab (CDF) and D0. On the way to record high luminosities, many issues related to the electromagnetic beam–beam interaction of colliding beams have been addressed. Below we present a short overview of the beam–beam effects in the Tevatron.

## BEAM–BEAM EFFECTS IN RUN I

For a detail history of the Tevatron accelerators, performance, and upgrades see Ref. [1]. In 1978 Fermilab decided that proton–antiproton collisions would be supported in the Tevatron, at a centre-of-mass energy of 1800 GeV, and that an Antiproton Source facility would be constructed to supply the flux of antiprotons needed for a design luminosity of $1 \times 10^{30}$ cm$^{-2}$ s$^{-1}$.

The Tevatron as a fixed target accelerator was completed in 1983. The Antiproton Source was completed in 1985 and the first collisions were observed in the Tevatron using operational elements of the CDF detector (then under construction) in October 1985. Initial operations of the collider for data-taking took place during a period from February to May 1987. A more extensive run took place between June 1988 and June 1989, representing the first sustained operation at the design luminosity. In this period of operation a total of 5 pb$^{-1}$ were delivered to CDF at 1800 GeV (centre-of-mass).

The initial operational goal of $1 \times 10^{30}$ cm$^{-2}$ s$^{-1}$ luminosity was exceeded during this run.

Collider Run I took place from August 1992 through February 1996 and employed six bunches in each beam on separated orbits (22 electrostatic separators aimed at mitigating the beam–beam limitations were installed by 1992). Antiproton source improvements supported an accumulation rate of $7 \times 10^{10}$ antiprotons h$^{-1}$. Run I ultimately delivered a total integrated luminosity of 180 pb$^{-1}$ to both the CDF and D0 experiments at $\sqrt{s} = 1800$ GeV. By the end of the run the typical luminosity at the beginning of a store was about $1.6 \times 10^{31}$ cm$^{-2}$ s$^{-1}$, a 60% increase over the Run I goal.

Even at the initial stages of the colliding beam operation, very high beam–beam tune shift parameters were achieved for both protons and antiprotons:

$$\xi = N_{IP} \frac{N_p r_p}{4\pi\varepsilon} \approx 0.018 - 0.025 \quad (1)$$

where $r_p$ denotes the classical proton radius, $N_p$ and $\varepsilon$ are the opposite bunch intensity and emittance, respectively, and $N_{IP} = 12$ was the total number of head-on collisions per turn with six-on-six bunches operation. It was realized that the beam–beam footprint covers almost all available tune space between the 3/7th and 2/5th resonances (see Fig. 1) (note that later, after installation of the low-beta insertions for Run I, the working point (WP) was established above half-integer); that the beam loss rates are strongly dependent on the tunes and $\xi$ (see Fig. 2); and the luminosity lifetime significantly deteriorates at the highest beam–beam parameters (see Fig. 3).

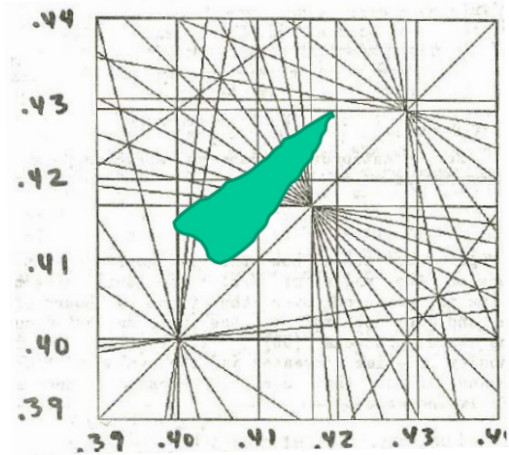

Figure 1: The Tevatron working point in Run I [2].

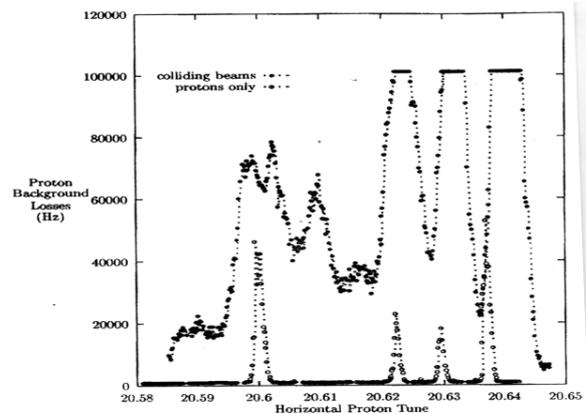

Figure 2: Proton background loss rate in the detectors vs. proton tune $Q_x$ with and without collisions [3].

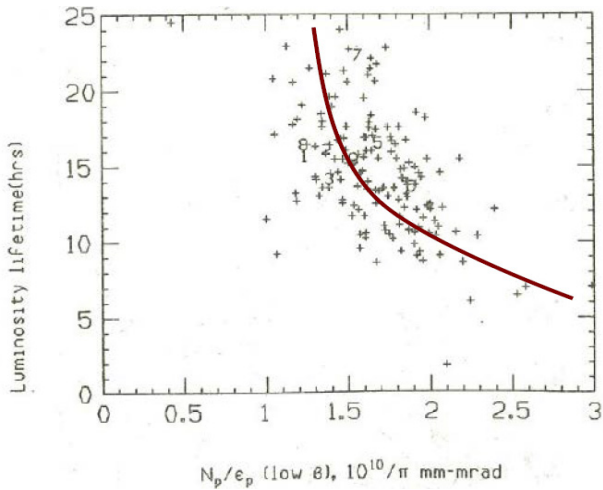

Figure 3: Initial luminosity lifetime as a function of proton brightness $N_p/\varepsilon_p$ [2].

Other effects observed were antiproton emittance blowup, transverse halo growth in antiprotons, and high losses in both beams [2, 3]. It was expected that in Collider Run II operation with 36 bunches in each beam, with a much higher antiproton intensity (and, consequently, luminosity) the antiproton helical orbits and tunes would vary significantly from bunch to bunch. The distribution of the antiproton tunes vs. longitudinal bunch position has been measured in a dedicated study and found to be in agreement with theory [4] (Fig. 4). The scale of the expected beam–beam effect was not very clear at that time, and as a safety measure the project for beam–beam compensation with electron lenses was started [5].

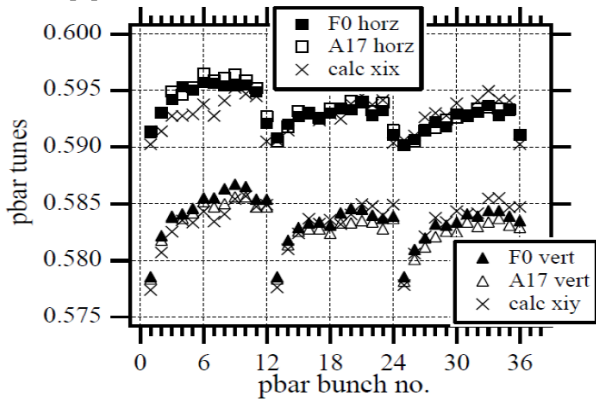

Figure 4: Measured and calculated antiproton tunes for colliding beam conditions (from Ref. [4]). Base pbar (antiproton) tunes were $Q_x/Q_y = 0.5855/0.5755$. A scale factor of ~0.65 was assumed for the tune shift from the head-on beam–beam interaction.

## BEAM–BEAM EFFECTS IN RUN II

For the most comprehensive reviews of the beam–beam effects in the Tevatron Collider Run II see Refs. [6, 7]. Collider Run II (2001 to 2011) had employed two new accelerators in the injector chain—the Main Injector (for a significant increase in the number of protons on the antiproton target and beam quality for the collider) and the Recycler (to provide storage for very large numbers of antiprotons—up to $6 \times 10^{12}$—and their cooling with stochastic and electron cooling systems). Four additional separators were installed to improve separation at the nearest parasitic crossings. At the end of Run II, typical Tevatron luminosities were well in excess of $3.4 \times 10^{32}$ cm$^{-2}$ s$^{-1}$, with record stores exceeding $4.3 \times 10^{32}$ cm$^{-2}$ s$^{-1}$ (or more than five times above the Run II goal) (see Table 1).

Table 1: Achieved performance parameters for Collider Runs I and II (typical values at the beginning of a store)

|  | 1988 to 1989 Run | Run Ib | Run II |
|---|---|---|---|
| Energy (com) (GeV) | 1800 | 1800 | 1960 |
| Protons/bunch ($\times 10^{10}$) | 7.0 | 23 | 29 |
| Antiprotons/bunch ($\times 10^{10}$) | 2.9 | 5.5 | 8.1 |
| Bunches/beam | 6 | 6 | 36 |
| Total antiprotons ($\times 10^{10}$) | 17 | 33 | 290 |
| P-emittance (rms, n) ($\pi$ μm) | 4.2 | 3.8 | 3.0 |
| Pbar emittance (rms, n) ($\pi$ μm) | 3 | 2.1 | 1.5 |
| $\beta^*$ (cm) | 55 | 35 | 28 |
| Luminosity (typical) ($10^{30}$ cm$^{-2}$ s$^{-1}$) | 1.6 | 16 | 350 |
| Luminosity integral (fb$^{-1}$) | $5 \cdot 10^{-3}$ | 0.18 | 11.9 |

During Collider Run II, beam losses during injection, ramp, and squeeze phases were mostly caused by the long-range beam–beam effects. Early in Run II, the combined beam losses in the Tevatron alone (the last accelerator out of a total of seven in the accelerator chain) claimed significantly more than half of the integrated luminosity (see Fig. 5). Due to various improvements, the losses have been reduced significantly down to some 20–30% in 2008 to 2009, paving the road to a many-fold increase of the luminosity. In 'proton-only' or 'antiproton-only' stores, the losses do not exceed 2–3% per specie. So, the remaining 8–10% proton loss and 2–3% antiproton loss is caused by beam–beam effects, as well as some 5–10% reduction of the luminosity lifetime through collision.

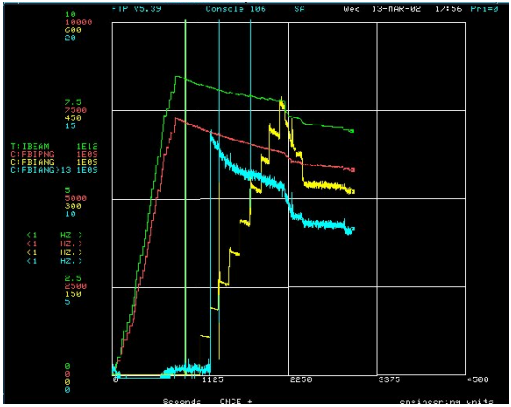

Figure 5: A typical plot of the collider 'shot' shows significant beam losses early in Run II (2002).

The particle losses for both beams *on the separated orbit* were driven by diffusion and exacerbated by limited longitudinal and/or transverse apertures. The intensity decay (see Fig.6) was well approximated by [6]:

$$\frac{\Delta N_{a,p}}{N_{a,p}} = 1 - \frac{N(t)}{N(t=0)} \propto \sqrt{t} \cdot \varepsilon_{a,p}^2 \frac{N_{p,a}}{\varepsilon_{p,a}} Q'^2_{a,p} \cdot F(\varepsilon_L, Q_{x,y}, S_{a-p}) \quad . \quad (2)$$

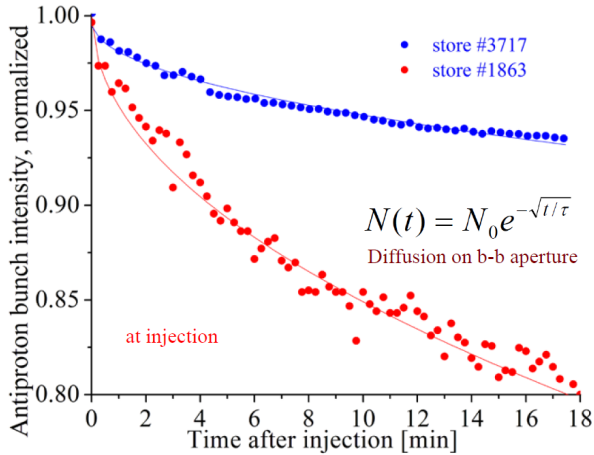

Figure 6: Decay of (normalized) intensity for antiproton bunch #1 at injection. The red dots are for store #1863 (16 October 2002) and the blue dots are store #3717 (8 August 2004). The blue and red lines represent fits according to $N(t) = N_0 e^{-\sqrt{t/\tau}}$ with parameters $N_0 = 32.5 \times 10^9$, $\tau = 7.4$ h and $N_0 = 55.7 \times 10^9$, $\tau = 69.8$ h, respectively [6].

The $\sqrt{t}$-dependence has also been observed for bunch length 'shaving' (slow reduction of the rms bunch length), while transverse emittances do not exhibit such dependence and usually either stay flat or grow slightly. Notably, the proton inefficiencies were higher than the antiproton ones, despite the factor of 3–5 higher proton intensity. That was due to significantly smaller antiproton emittances (see Eq. (2)). Due to the strong dependence of the losses on the chromaticities $Q'_{x,y}$ and beam separation:

$$S = \sqrt{(\Delta x/\sigma_{x\beta})^2 + (\Delta y/\sigma_{y\beta})^2} \quad . \quad (3)$$

special measures were taken to reduce the former (octupoles and feedback systems allowed $Q'$ to decrease to almost zero) and increase the latter via the increase of limiting physical apertures followed by the increase of the helix size and/or optimization of the HV separator voltages (see Fig. 7).

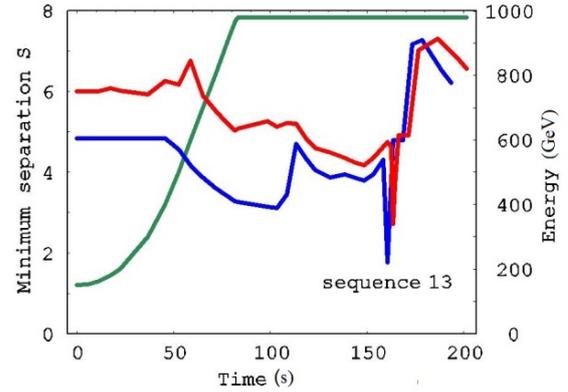

Figure 7: Minimum radial separation (see Eq. (3)) on ramp and during the low-beta squeeze. The green line represents the beam energy on the ramp. The blue and red lines represent $S(t)$ for the helix configurations used ca. January 2002 and August 2004, respectively (from Ref. [6]).

The *head-on beam–beam effects* during the colliding beams stores had been significantly amplified by the presence of the parasitic long-range interactions and unequal beam sizes at the main interaction points (IPs). They were characterized by the record high beam–beam parameters for both protons and antiprotons (the head-on tune shifts up to about $\xi = 0.020$–$0.025$ for both protons and antiprotons (see Fig. 8), in addition to the long-range tune shifts of $\Delta Q^p = 0.003$ and $\Delta Q^a = 0.006$, respectively (see Fig. 9)), and remarkable differences in the beam dynamics of individual bunches.

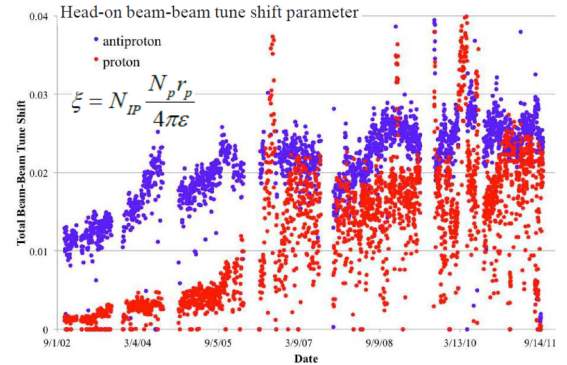

Figure 8: Proton (red) and antiproton (blue) head-on beam–beam tune shifts early in high energy physics (HEP) stores calculated from the measured beam parameters from 2002 to 2011.

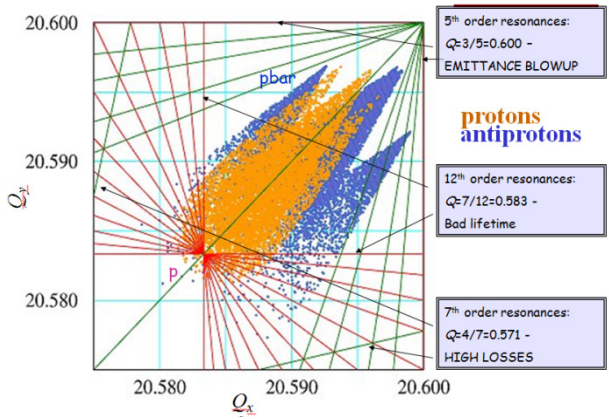

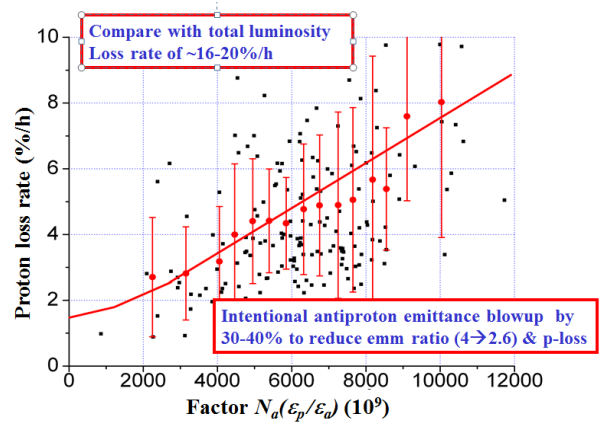

Figure 9: Tevatron proton and antiproton tune distributions superimposed onto a resonance line plot. The red and green lines are various sum and difference tune resonances of up to twelfth order. The blue dots represent calculation of the tune distributions for all 36 antiproton bunches; yellow dots represent the protons. The tune spread for each bunch is calculated for particles up to 6σ amplitude taking into account the *measured* intensities and emittances (from Ref. [6]). The most detrimental effects occur when particle tunes approach the resonances. For example, an emittance growth of the core of the beam is observed near the fifth-order resonances (defined as $nQ_x + mQ_y = 5$, such as $Q_{x,y} = 3/5 = 0.6$) or fast halo particle loss near twelfth-order resonances (for example, $Q_{x,y} = 7/12 \approx 0.583$).

The proton loss rate was also strongly affected by transverse size mismatch for head-on collisions of larger size proton bunches with smaller size antiproton bunches (see Fig. 10). Our studies of the phenomenon in 2003 to 2005 can be summarized as [6]:

$$\frac{1}{\tau_p} = \frac{1}{N_p}\frac{dN_p}{dt} \propto N_a \cdot \left(\frac{\varepsilon_p}{\varepsilon_a}\right)^2 F_2(Q_{x,y}, Q', Q'', M) \quad . \quad (4)$$

where $M$ stands for bunch position in the bunch train (see Fig. 11). In order to avoid a large emittance ratio $\varepsilon_p/\varepsilon_a$, the antiproton emittances are routinely diffused at the beginning of HEP stores by a wide band transverse noise to a directional strip line, so the ratio is kept to about 2.6–3.

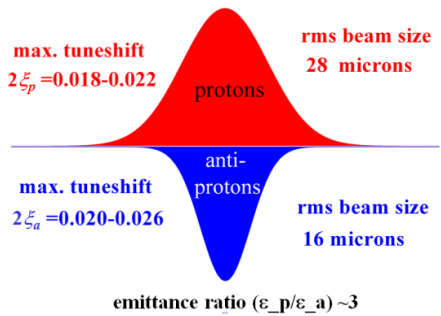

Figure 10: Mismatch of the proton and antiproton transverse beam sizes at the Tevatron IPs.

Figure 21: Measured proton loss rates at the beginning of HEP stores vs. factor $N_a(\varepsilon_p/\varepsilon_a)$.

The factor $F$ in Eq. (4) is to indicate strong dependence of the losses on the second order betatron tune chromaticity $Q'' = d^2Q/d(\Delta p/p)^2$. Numerical modelling [7]—that was later confirmed by experiments—showed that the deterioration of the proton lifetime was caused by a decrease of the dynamical aperture for off-momentum particles due to head-on collisions (Fig. 12).

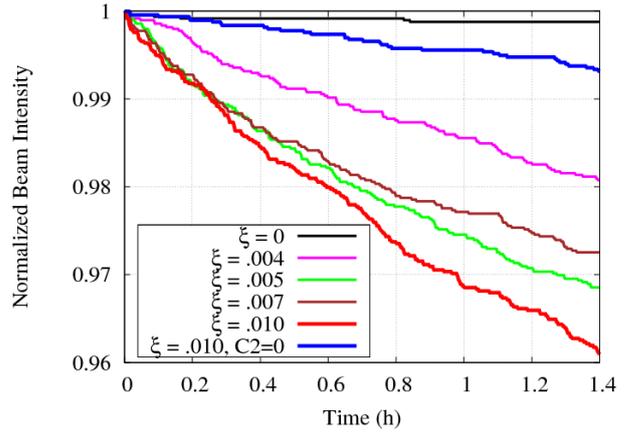

Figure 12: Proton intensity evolution for different values of the beam–beam tune shift parameter per IP from $\xi = 0$ to $\xi = 0.01$; without and with compensation of the chromaticity of $\beta^*$ ($C2 = 0$) (from Ref. [7]).

It was discovered that the Tevatron optics had large chromatic perturbations, e.g. the value of $\beta^*$ for off-momentum particles could differ from that of the reference particle by as much as 20%. Also, the high value of second-order betatron tune chromaticity $Q'' = d^2Q/d(\Delta p/p)^2$ generated a tune spread of ~0.002. A rearrangement of the sextupole magnet circuits in order to correct the second order chromaticity was planned and implemented before the 2007 shutdown and led to some 10% increase in the luminosity integral/store for 2009 operations (see Fig. 13).

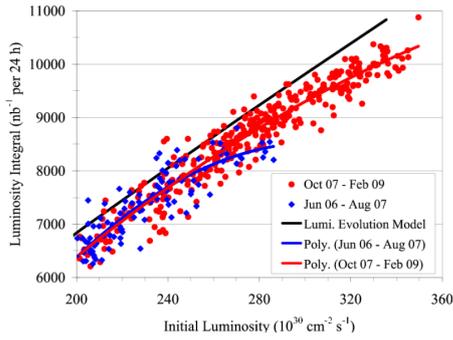

Figure 13: Luminosity integral normalized for 24 h store vs. initial luminosity. Blue points and curve, before the second-order chromaticity correction; red, after the correction. The black line represents the maximum possible luminosity integral for the given beam parameters in the absence of beam–beam effects [7].

The collider luminosity lifetime is determined by the speed of the emittance growth, beam intensity loss rates and bunch lengthening (which affects hour-glass factor $H$):

$$\tau_L^{-1} = \frac{dL(t)}{L(t)dt} = |\tau_\varepsilon^{-1}| + \tau_{Na}^{-1} + \tau_{Np}^{-1} + \tau_H^{-1}. \quad (5)$$

At the end of Tevatron Collider Run II, the luminosity loss rates were in the range 19–21%/h at the beginning of storage. For the 2010 to 2011 HEP stores with a range of initial luminosities between 3.0 and $4.3 \times 10^{32}$ cm$^{-2}$ s$^{-1}$, the largest contribution to luminosity decay came from beam emittance growth with a typical time of $\tau_\varepsilon \sim$ 9–11 h. The growth is dominated by intrabeam scattering (IBS) in the proton bunches, with small contributions from the IBS in antiprotons and external noises. Beam–beam effects, if noticeable, usually manifest themselves in reduction of the beam emittances or their growth rates rather than in increases.

The antiproton bunch intensity lifetime $\tau_a \sim$ 16–18 h is dominated by the luminosity burn rate, which accounts for 80–85% of the lifetime, while the remaining 10–15% comes from parasitic beam–beam interactions with protons. Proton intensity loss varies in a wide range $\tau_p \sim$ 25–45 h and is driven mostly (~50%) by the head-on beam–beam interactions with smaller size antiprotons at the main IPs. The proton lifetime caused by inelastic interactions with antiprotons in collisions and with residual gas molecules varies from 300–400 h. The hourglass factor decays with $\tau_H \sim$ 70–80 h due to the IBS, again, mostly in proton bunches. Beam–beam effects may lead to reduction of the proton bunch length growth (longitudinal 'shaving') in a poorly tuned machine. Combining all of these loss rates together, one can estimate the hit on the luminosity lifetime $\tau_L$ due to the beam–beam effects as 12–17% (which is equal to (2.5–3.5%/h)/(19–21%/h)).

The luminosity integral $I = \int L dt$—the sole critical parameter for HEP experiments—depends on the product of peak luminosity and the luminosity lifetime, e.g. for a single store with initial luminosity $L_0$ and duration $T \sim 16$ h, the integral is $I \approx L_0 \tau_L \ln(1 + T/\tau_L)$. Therefore, the full impact of the beam–beam effects on the luminosity integral should include beam–beam driven proton and antiproton losses at the injection energy (about 5% and 1%, respectively), on the energy ramp (2% and 3%), and in the low-beta squeeze (1–2% and 0.5%), which proportionally reduce the initial luminosity $L_0$. So, altogether, at the last operational stage of the Tevatron collider present, the beam–beam effects reduce the luminosity integral by 23–33% (see Fig. 14).

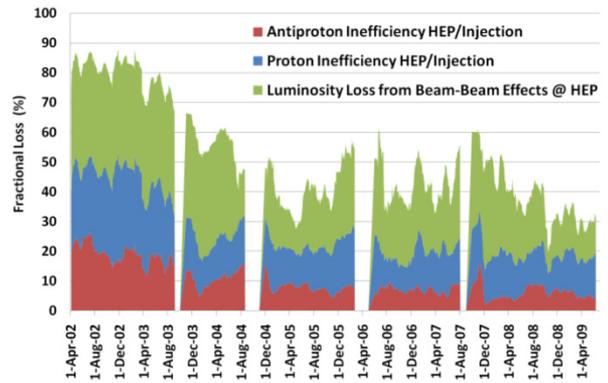

Figure 14: Evolution of beam losses in 2002 to 2009. the red band shows fractional losses of antiprotons between injection into the Tevatron and the start of collisions; the blue band is for the loss of protons; the green band is for the fractional reduction of the luminosity integral caused by beam–beam effects in collisions (from Ref. [8]).

The Tevatron Collider performance history analysis [9] shows that the luminosity increases occurred after numerous improvements; some were implemented during operation and others were introduced during regular shutdown periods. The actions that helped us to keep the beam–beam effects under control during the Run II operations included: i) *at injection, ramp and low-beta squeeze*: opened apertures (replaced magnets), increased helix size $S$, chromaticity $Q'$ reduction (with help of the transverse dampers and octupoles), optimization of the helices (many times), improved emittances from the injectors; ii) *at low beta (in collision stores)*: the use of additional separators, helix optimization (increased separation at the first long-range IPs), reduction of the chromaticity $Q'$, pbar transverse blowup, tune stabilization; reduction of the chromatism of beta-function ($Q''$); iii) *trustable beam-beam simulations*; iv) *operational machine stabilization*: stable (repeatable) intensities and emittances from injectors, drastically stabilized Tevatron; v) *outstanding development of the beam diagnostics*: there were three cross-calibrated instruments for the tune measurements, three types of emittance monitors, three intensity monitors, two luminosity monitors, and several types of beam position monitors [10].

## BEAM–BEAM COMPENSATION, TELS

Detailed description of the Tevatron Electron Lenses (TELs) and results of the beam–beam compensation studies can be found in Refs. [5, 11–13]. Electron lenses (e-lenses) employ electromagnetic fields of strongly magnetized high-intensity electron beams and were originally proposed for the compensation of head-on and long-range beam–beam effects in the Tevatron [5] (see Fig. 15). The lens employs a low-energy beam of electrons that collides with the high-energy proton or antiproton bunches over an extended length. Electron space charge forces are linear at distances smaller than the characteristic beam radius $r < a_e$ but scale as $1/r$ for $r > a_e$. Correspondingly, such a lens can be used for linear long-range beam–beam and nonlinear head-on beam–beam force compensation depending on the beam-size ratio $a_e/\sigma$ and the current-density distribution $j_e(r)$.

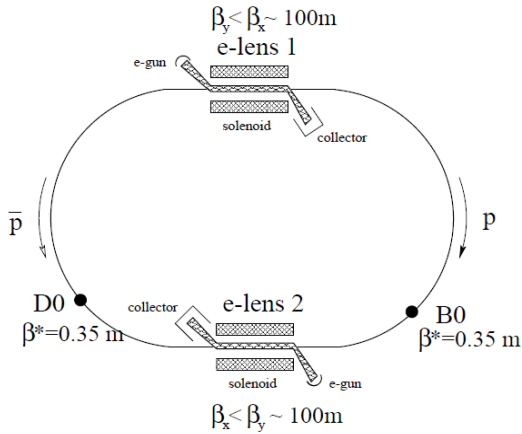

Figure 15: Schematic Tevatron layout with two electron lenses [5].

Main advantages of the e-lenses are: i) the electron beam acts on high-energy beams only through EM forces, with no nuclear interactions; ii) fresh electrons interact with the high-energy particles each turn, leaving no possibility for coherent instabilities; iii) the electron current profile (and, thus, the EM field profiles) can easily be changed for different applications (see Fig. 16); iv) the electron-beam current can be quickly varied, e.g. on a timescale of bunch spacing in accelerators (see Fig. 17).

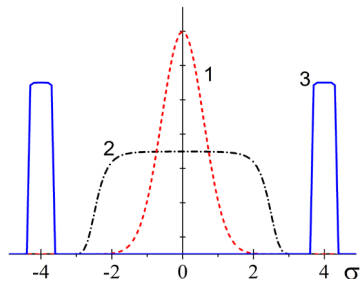

Figure 16: The transverse electron current profiles in electron lenses for: (1) space charge and head-on beam-beam compensation; (2) for bunch-by-bunch tune spread compensation; (3) halo collimation.

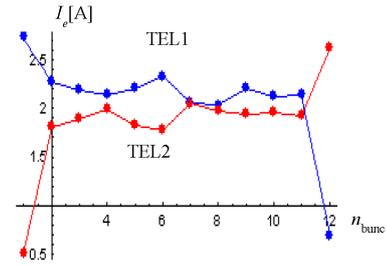

Figure 17: Variation of the currents in two electron lenses needed for long-range tune shift compensation in the Tevatron (bunch spacing is 396 s$^{-9}$).

Two electron lenses were built and installed in the A11 and F48 locations of the Tevatron ring. They use a 1–3 A, 6–10 kV e-beam generated at the 10–15 mm diameter thermionic cathodes immersed in a 0.3 T longitudinal magnetic field and aligned onto the (anti)proton beam orbit over ~2 m length inside a 6 T SC solenoid [11] (see Fig. 18).

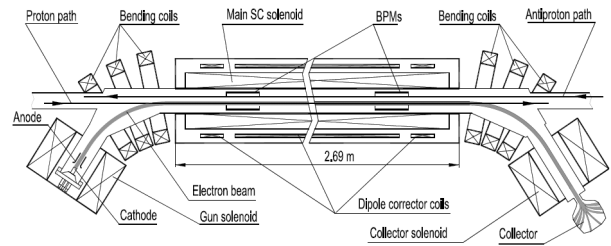

Figure 18: General layout of the Tevatron Electron Lens.

The high-energy protons are focused by the TEL and experience a positive betatron tune shift:

$$dQ_{x,y} = +\frac{\beta_{x,y} L_e r_p}{2\gamma ec} \cdot j_e \cdot \left(\frac{1-\beta_e}{\beta_e}\right) \quad (6)$$

In the long-range beam–beam compensation (BBC) experiments [12], a large-radius electron beam was generated $a_e \approx 3\sigma$, therefore the tune shift was about the same for most protons in the bunch. The tune shift for the antiprotons is of about the same magnitude, but negative. Maximum measured tune shift for 980 GeV protons was about 0.01 (see Fig. 19).

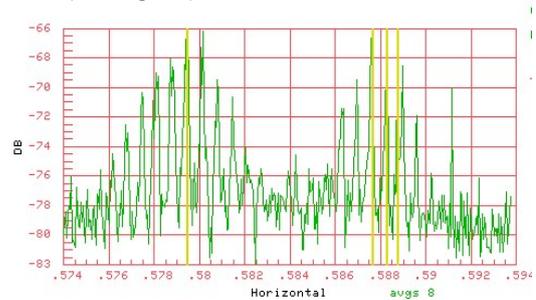

Figure 19: Horizontal tune shift of the 980 GeV proton beam induced by TEL-1 [13].

In the BBC demonstration experiment [12], the electron beam of TEL-2, which was installed at the A11 location with a large vertical beta-function of $\beta_y = 150$ m, was centred and timed onto bunch #12 without affecting any other bunches. When the TEL peak current was increased to $J_e = 0.6$ A, the lifetime $\tau = N/(dN/dt)$ of bunch #12 went from about 12 h up to 26.6 h (see Fig. 20.)

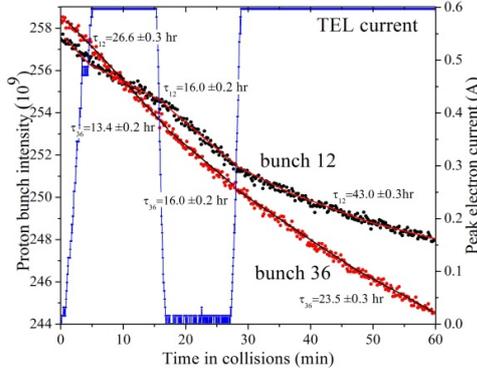

Figure 20: Intensities of proton bunches #12 and #36 early in store #5119 with $L_0 = 1.6 \times 10^{32}$ cm$^{-2}$ s$^{-1}$ (see the text) [12, 13].

At the same time, the lifetime of bunch #36, an equivalent bunch in the third bunch train, remained low and did not change significantly (at 13.4 h lifetime). When the TEL current was turned off for fifteen minutes, the lifetimes of both bunches were, as expected, nearly identical (16 h). The TEL was then turned on again, and once again the lifetime for bunch #12 improved significantly to 43 h while bunch #36 stayed poor at 23.5 h. This experiment demonstrates a factor of two improvement in the proton lifetime due to compensation of beam–beam effects with the TEL.

Proton lifetime, dominated by beam–beam effects, gradually improves and reaches roughly 100 h after 6 h to 8 h of collisions; this is explained by a decrease in antiproton population and an increase in antiproton emittance, both contributing to a reduction of the proton beam–beam parameter. To study the effectiveness of BBC later in the store, the TEL was repeatedly turned on and off every half hour for 16 h, again on bunch #12. The relative improvement $R$, defined as the ratio of the proton lifetime with the TEL and without, dropped from $R = 2.03$ to $R = 1.0$ in about 10 h. At this point, the beam–beam effects have become very small, providing little to compensate. Similar experiments in several other stores with initial luminosities ranging from $L_0 = 1.5 \times 10^{32}$ cm$^{-2}$ s$^{-1}$ to $2.5 \times 10^{32}$ cm$^{-2}$ s$^{-1}$ repeated these results [13].

Results of many experiments with TELs are reported in Ref. [13], and studies of non-linear BBC with a Gaussian electron beam current profile are presented in Ref. [14]. TELs were not used routinely for the BBC in the Tevatron because beam–beam losses were effectively controlled by other means, as described above. Numerical simulations [15] predict a beneficial effect of electron lenses upon the ultimate intensity LHC beam lifetime.

Besides the BB compensation, the TELs were used for operational abort gap beam removal [16] (see Fig. 21) and for beam halo collimation [17] (see Fig. 22).

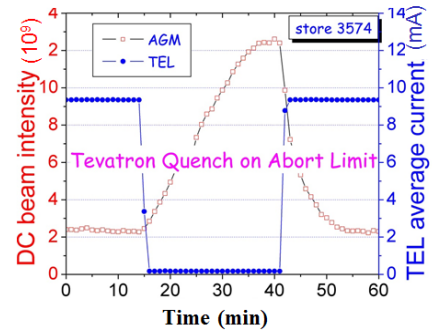

Figure 21: Effective removal of the DC beam from the Tevatron abort gaps by TEL [16].

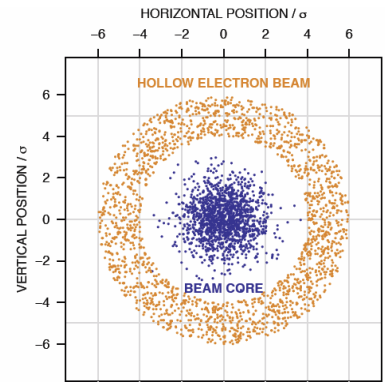

Figure 22: Geometry of the beams for the transverse beam halo removal experiment in the Tevatron [17].

## SUMMARY

The beam–beam effects in the Tevatron turned from 'tolerable' (in Collider Run I) to 'very detrimental' (early Collider Run II). We experienced a broad variety of the effects—in both beams, at all stages of the cycle, long-range, and head-on. The Tevatron team has been able to address them and provide critical contribution to a more than 30-fold luminosity increase by the end of Run II (see Fig. 23). We have also enriched beam physics by experimental studies, development of advanced theory, and trustable modelling tools to simulate the beam–beam effects, development of the electron lenses, and the first demonstration of active beam–beam compensation.

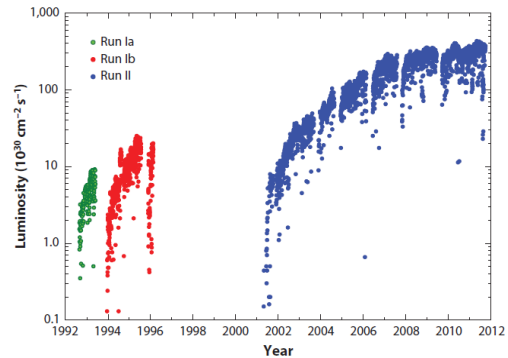

Figure 23: Initial luminosity for all collider stores [1].

The Tevatron collider program ended on 30 September 2011. The machine has worked extremely well for 25 years and has greatly advanced accelerator technology and beam physics. Its success is a great tribute to the Fermilab staff. Among those who contributed to the exploration of the beam–beam effects in the Tevatron were: (*Tevatron Collider Run I*) J. Annala, S. Assadi, P. Bagley, D. Finley, G. Goderre, D.A. Herrup, R. Johnson, J. Johnstone, E. Malamud, M. Martens, L. Michelotti, S. Mishra, G. Jackson, S. Peggs, S. Saritepe, D. Siergiej, P. Zhang; (*Beam–beam effects in the Collider Run II*) Yu. Alexahin, J. Annala, D. Bollinger, V. Boocha, J. Ellison, B. Erdelyi, N. Gelfand, B. Hanna, H.J. Kim, P. Ivanov, A. Jansson, A. Kabel, V. Lebedev, P. Lebrun, M. Martens, R.S. Moore, V. Nagaslaev, R. Pasquinelli, V. Sajaev, T. Sen, E. Stern, D. Shatilov, V. Shiltsev, G. Stancari, D. Still, M. Syphers, A. Tollestrup, A. Valishev, M. Xiao; (*Beam–beam compensation*) A. Aleksandrov, Y. Alexahin, L. Arapov, K. Bishofberger, A. Burov, C. Crawford, V. Danilov, B. Fellenz, D. Finley R. Hively, V. Kamerdzhiev, S. Kozub, M. Kufer, G. Kuznetsov, P. Logatchov, A. Martinez, F. Niell, M. Olson, V. Parkhomchuk, H. Pfeffer, V. Reva, G. Saewert, V. Scarpine, A. Seryi, A. Shemyakin, V. Shiltsev, N. Solyak, G. Stancari, B. Sukhina, V. Sytnik, M. Tiunov, L. Tkachnko, A. Valishev, D. Wildman, D. Wolff, X. Zhang, F. Zimmermann, A. Zinchenko. All the credits for the fascinating results presented above in this review should go to these dedicated researchers. Eight further related presentations have been made at this Workshop [18–25].